\documentclass[epj]{svjour}

\usepackage{graphicx}
\usepackage{fancyhdr}
\usepackage{amsmath}
\usepackage{amssymb}
\def\makeheadbox{{%  remove EPJC header
 }}
\newcommand{\tb}{\tan \beta}
\def\mytitle{My title} 
\def\myauthors{My name}  
\def\mytype{My type of session}
\def\mysession{My session}
\def\mytitle{Constraints on mSUGRA} %Put your title here!
\def\myauthors{M. Mondrag\'on}    %Put your name here!
\def\mytype{Contributed Talk}    
\def\mysession{Cosmology and Astrophysics}
\begin{document}
\title{Constraints on mSUGRA through entropy and abundance criteria}
\author{L.G. Cabral-Rosetti\inst{1}
 \and
 M. Mondrag\'on\inst{2}% etc
\thanks{\emph{Email:} myriam@fisica.unam.mx}%
 \and 
 L. Nellen\inst{3}
 \and
 D. N\'u\~nez\inst{3}
 \and
 R. Sussmann\inst{4}\thanks{On sabbatical leave from Inst. de Ciencias
   Nucleares, UNAM} 
 \and
 J. Zavala\inst{3}
}                     % Do not remove
\institute{Depto. de Posgrado,
Centro Interdisciplinario de Investigaci\'on\\ y
Docencia en Educaci\'on T\'ecnica (CIIDET), Qro., M\'exico
\and Inst. de F\'\i sica, Universidad Nacional Aut\'onoma de M\'exico (IF-UNAM), M\'exico
\and Inst. de Ciencias Nucleares,  Universidad Nacional Aut\'onoma de
M\'exico (ICN-UNAM), M\'exico 
\and Inst. de F\'\i sica, Universidad de Guanajuato (IFUG),  M\'exico
}
\date{}
\abstract{ We derive an expression for the entropy of a present dark
  matter halo described by a Navarro-Frenk-White modified model with a
  central core.
  The comparison of this entropy with the one of the halo at the
  freeze-out era allows us to obtain an expression for the relic
  abundance of neutralinos, which in turn is used to constrain the
  parameter space in mSUGRA models, when used with the WMAP
  observations.  Moreover, by joning these results with the ones
  obtained from the usual abundance criteria, we are able to clearly
  discriminate validity regions among $\tb$ values of the mSUGRA
  model, by demanding both criteria to be consistent with the 2 sigma
  bounds of the WMAP observations for the relic density:
  $0.112<\Omega h^2<0.122$.  We found that for 
  $sgn~ \mu =+$, small values of $\tb$ are not favored; only for $\tb
  \sim 50$ are both criteria significantly consistent. The use of both
  criteria also allows us to put a lower bound on the neutralino mass,
  $m_{\chi}\geq151$GeV.
}
\PACS{
      {14.80.Ly}{Supersymmetric partners of known particles}   \and
      {95.35.+d}{Dark matter} \and
      {98.62.Gq}{Galactic halos}
     } % end of PACS codes
%end of abstract
%
\maketitle
%DO NOT REMOVE THIS LINE
%

\section{Introduction}

Supersymmetry models which have the neutralino as the lightest
supersymmetric particle (LSP) and as a candidate for dark matter (DM),
have several parameters that can be constrained by the bounds on the
present density of DM, $\Omega_{CDM}$, that come from several outstanding
observations such as the Cosmic Microwave Background radiation (CMBR)
\cite{WMAP}, Galaxy clustering, Supernovae and Lyman $\alpha$ forest. One
of the most recent works which combines all these data leads to:
$0.112\leq\Omega_{CDM}h^2\leq0.122$ \cite{seljak}.  In particular for mSUGRA
models these constraints have been obtained using the standard approach
\cite{belanger,constraint}, which is based in the Boltzmann equation
considering that after the ``freeze-out'' era, neutralinos cease to
annihilate keeping its number constant. In such an approach, the relic
density of neutralinos is approximately: $\Omega_{\chi}\approx 1/\langle\sigma v\rangle$, where
$\langle\sigma v\rangle$ is the thermally averaged cross section times the relative
velocity of the LSP annihilation pair. Within the mSUGRA model five
parameters ($m_0$, $m_{1/2}$, $A_0$, tan$\beta$ and the sign of $\mu$) are
needed to specify the supersymmetric spectrum of particles and the
final relic density. We will use the numerical code micrOMEGAs
\cite{micro} to compute the relic density following the past scheme
which will be called the ``abundance criterion'' (AC).

Just after ``freeze-out'', we can consider neutralinos then as forming a 
Maxwell-Boltzmann (MB) gas in thermal equilibrium with other components of
the primordial cosmic structures. In the present time, such a gas is
almost colisionless and either constitutes galactic halos and larger structures
or it is in the process of its formation. In this context, we can conceive
two equilibrium states for the neutralino gas, the decoupling (or 
``freeze-out'') epoch and its present state as a virialized system. Computing
the entropy per particle for each one of this states we can use an 
``entropy consistency'' criterion (EC) using theoretical and empirical 
estimates for this entropy to obtain the relic density of neutralinos 
($\Omega_{\chi}$).

Our objective is then to use AC and EC criteria, to obtain constraints
for the parameters of the mSUGRA model by demanding that both criteria
must be consistent with each other and with the observational
constraints of $\Omega_{CDM}$.

\section{Abundance criterion}

Relic abundance of some stable SUSY species $\chi$ is defined as 
$\Omega_{\chi}=\rho_{\chi}/\rho_{crit}$, where $\rho_{\chi}=m_{\chi}n_{\chi}$
is the relic's mass density ($n_{\chi}$ is the number density), $\rho_{crit}$
is the critical density of the Universe (see \cite{kamion} for a review
on the standard method to compute the relic density). The time evolution of
$n_{\chi}$ is given by the Boltzmann equation:
\begin{equation}\label{bolt}
\frac{dn_{\chi}}{dt}=-3Hn_{\chi}-\langle\sigma v\rangle(n_{\chi}^2
-(n_{\chi}^{eq})^2)
\end{equation}
where $H$ is the Hubble expansion rate, $\langle\sigma v\rangle$ is the thermally
averaged cross section times the relative velocity of the LSP
annihilation pair and $n_{\chi}^{eq}$ is the number density that species
would have in thermal equilibrium. In the early Universe, the
neutralinos ($\chi$) were initially in thermal equilibrium,
$n_{\chi}=n_{\chi}^{eq}$. As the Universe expanded, their typical
interaction rate started to diminish an the process of annihilation
froze out. Since then, the number of neutralinos in a comoving volume
has remained basically constant.

There are several ways to solve equation (\ref{bolt}), one of the more
used is based on the ``freeze-out'' approximation (see for example
\cite{gondolo}).  However in order to have more precision, we will use
the exact solution to Boltzmann equation using the public numerical
code micrOMEGAs 1.3.6 \cite{micro} which calculates the relic density
of the LSP in the Minimal Supersymmetric Standard Model (MSSM). We
will take and mSUGRA model and its five parametersas input parameters
for micrOMEGAs and use {\it Suspect} \cite{suspect}, which comes as an
interface to micrOMEGAs, to calculate the supersymmetric mass spectrum.

Using micrOMEGAs, we can obtain the relic density for any region of the 
parameter space to discriminate regions that are consistent with the WMAP 
constraints in this abundance criterion. 

\section{Entropy consistency criterion}

Since the usual MB statistics that can be formally applied to the
neutralino gas at the ``freeze-out'' era can not be used to describe
present day neutralinos subject to a long range gravitational
interaction making up non-extensive systems, it is necessary to use
the appropriate approach that follows from the microcanonical ensemble
in the ``mean field'' approximation, which yields an entropy definition
that is well defined for a self-gravitating gas in an intermediate
state. Such an approach is valid at both the initial (``freeze-out''
era, $f$) and final (virialized halo structures, $h$) states that we
wish to compare. Under these conditions, the change in the entropy per
particle ($s$) between these two states is given by \cite{Cabral-Rosetti:2004kd}:
\begin{equation}\label{entropy}
  s^h-s^f=ln\left[\frac{n_{\chi}^f}{n_{\chi}^h}\left(\frac{x^f}{x^h}\right)^{3/2}\right]~,
\end{equation}
where $x=m_{\chi}/T$, $T$ is the temperature of the gas. A region that
fits with the conditions associated with the intermediate scale is the
central region of halos ($~10 pc^3$ within the halo core); evaluating
the thermodynamical quantities at this region, using equation
(\ref{entropy}) and some extra assumptions (conservation of photon
entropy), it is possible to construct a theoretical estimate for $s^h$
that depends on the nature of neutralinos ($m_{\chi}$ and $\langle\sigma v\rangle$),
initial conditions (given by $x^f$), cosmological parameters
($\Omega_{\chi}$, the Hubble parameter, $h$) and structural parameters of
the virialized halo (central values for temperature and density); for
details of these and the following, see section IV of
\cite{Cabral-Rosetti:2004kd}.

An alternative estimate for $s^h$ can be made based on empirical
quantities for observed structures in the present Universe using the
microcanonical entropy definition in terms of phase space volume, but
restricting this volume to the actual range of velocities accessible
to the central particles. That is, restricting the escape velocity up
to a maximal value $v_e(0)$ which is related to the central velocity
dispersion of the halo ($\sigma_h$) by an intrinsic parameter $\alpha$:
$v_e^2(0)\sim\alpha\sigma_h^2(0)$. In a  recent work 
\cite{nuevo}, we estimate the value of $\alpha$ using an NFW modified
model with a central core, and obtain $16.4<\alpha <27.8$. The range of
values allowed for this parameter is of the highest importance to
determine the allowed region of the parameter space in the mSUGRA
model as will be clear in the results presented on next section.

Equating the theoretical an empirical estimates for the entropy per
particle it is obtained a relation for the relic abundance of
neutralinos using the EC criterion\footnote{This formula is a small
  modification to the one presented in \cite{Cabral-Rosetti:2004kd}}:
\begin{equation}\label{consistency}
ln(\Omega_{\chi}h^2)=10.853-x^f+ln\left[\frac{(x^f\alpha)^{3/2}m_{\chi}}{f_g^*(x^f)}\right]
\end{equation}
where $f_g^*(x^f)$ is a function related to the degrees of freedom at
the ``freeze-out'' time (see for example \cite{gondolo}) that will be
described elsewhere \cite{nuevo}.

Modifying the program micrOMEGAs, we obtained the value for $x^f$ for
any region of the parameter space and then $\Omega_{\chi}$ using
Eq.~(\ref{consistency}), therefore we were  able to discriminate
regions that are consistent with the WMAP constraints for the EC
criterion.

\section{Results and Conclusions}

Using both the AC and EC that have been described in the preceeding
sections, we can compute the total mass density of neutralinos present
today and constrain the region in the mSUGRA parameter space where
both criteria are fulfilled. Out of the five parameters, we will
fix $\mu >0$.
Then our strategy is to explore wide regions for the values of 
the other four parameters, by means of a bi-dimensional analysis
in the $m_0 - m_{1/2}$ plane for different fixed values
of $A_0$ and tan$\beta$. It is important to mention that we are not
presenting an exhaustive search in all the possible regions, but
we concentrated on those regions which have received more attention
in the literature, see for example \cite{belanger}.

\begin{figure}\centering
\includegraphics[height=15cm, width=4.8cm, viewport=200 0 300 350]{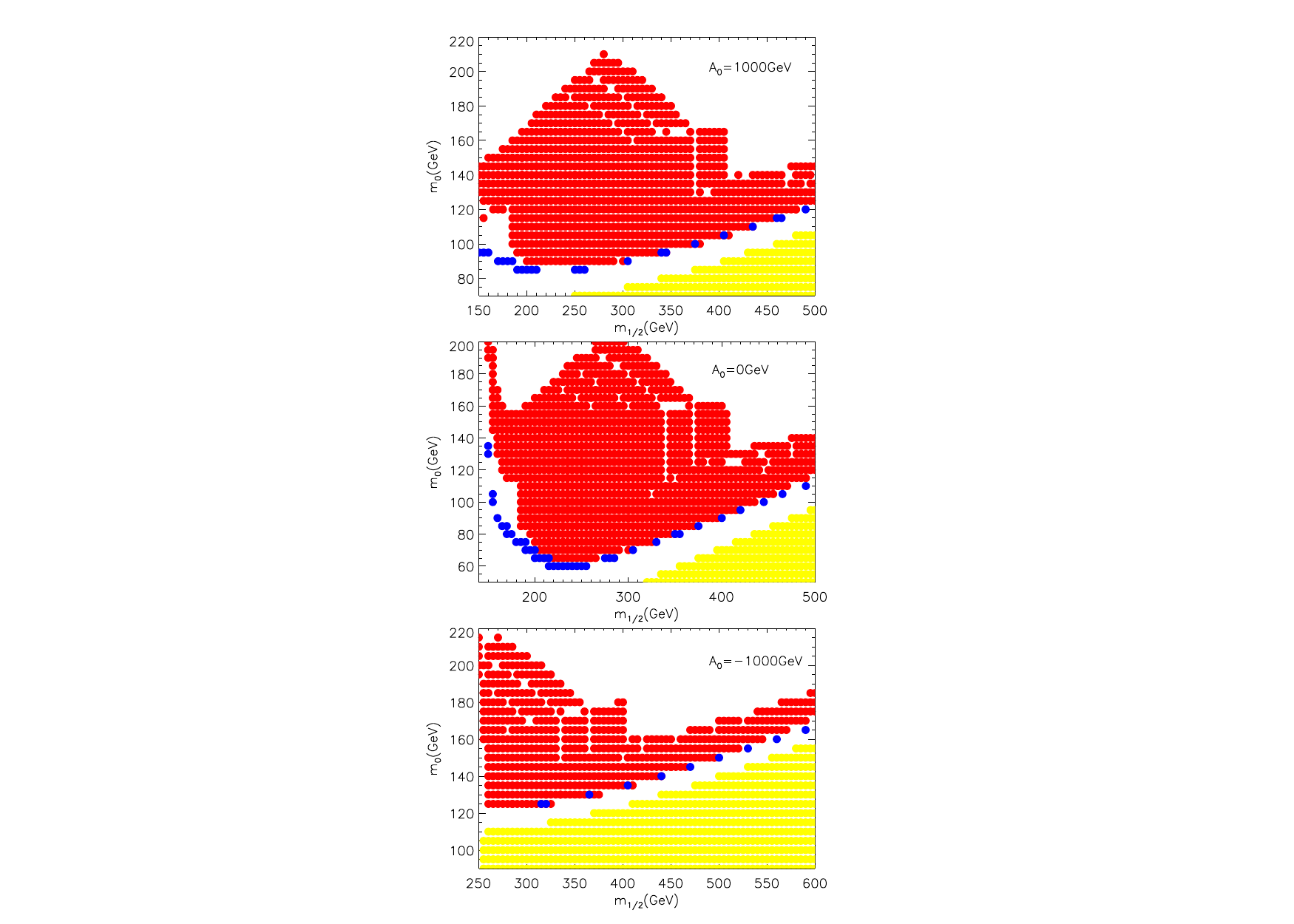}
\caption{Allowed regions in the parameter space for AC (lighter
  gray/red) and EC (darker grey/blue) for the mSUGRA model with
  sgn$\mu=+$, tan$\beta=10$, and $A_0=1000$ GeV, top panel, $A_0=0$ GeV,
  middle panel, and $A_0=-1000$ GeV, bottom panel. The figures show
  the so called bulk and coannihilation regions. The yellow region
  shows where the stau is the LSP.}
\label{tb10new}
\end{figure}

In Fig.~(\ref{tb10new}), we present the results for tan$\beta=10$, for
three values of $A_0$, namely $A_0=1000, 0, -1000$ GeV, shown in the
top, middle and bottom panels respectiveley.  The yellow region (lower
right corner) is where the $\tilde{\tau}$ is the LSP, the lighter and
darker areas (red and blue for the online version in colours) define
the allowed regions for the EC and AC respectively according to the
observed DM density. The area of the EC region depends on the size of
the interval of values of the parameter $\alpha$,  the
lower and upper bounds of $\alpha$ determine the upper and lower
boundaries of the EC region. As can be seen from the figure, the
region where both criteria are fullfilled is very small, in fact, only
for the highest values of $\alpha$ there is an intersection between both
criteria. This behavior holds for all values of $A_0$ in the interval
$[-1000, 1000]$ GeV, here we are showing only the extreme and central
values.

\begin{figure}\centering
\includegraphics[height=15cm, width=4.8cm, viewport=200 0 300 350]{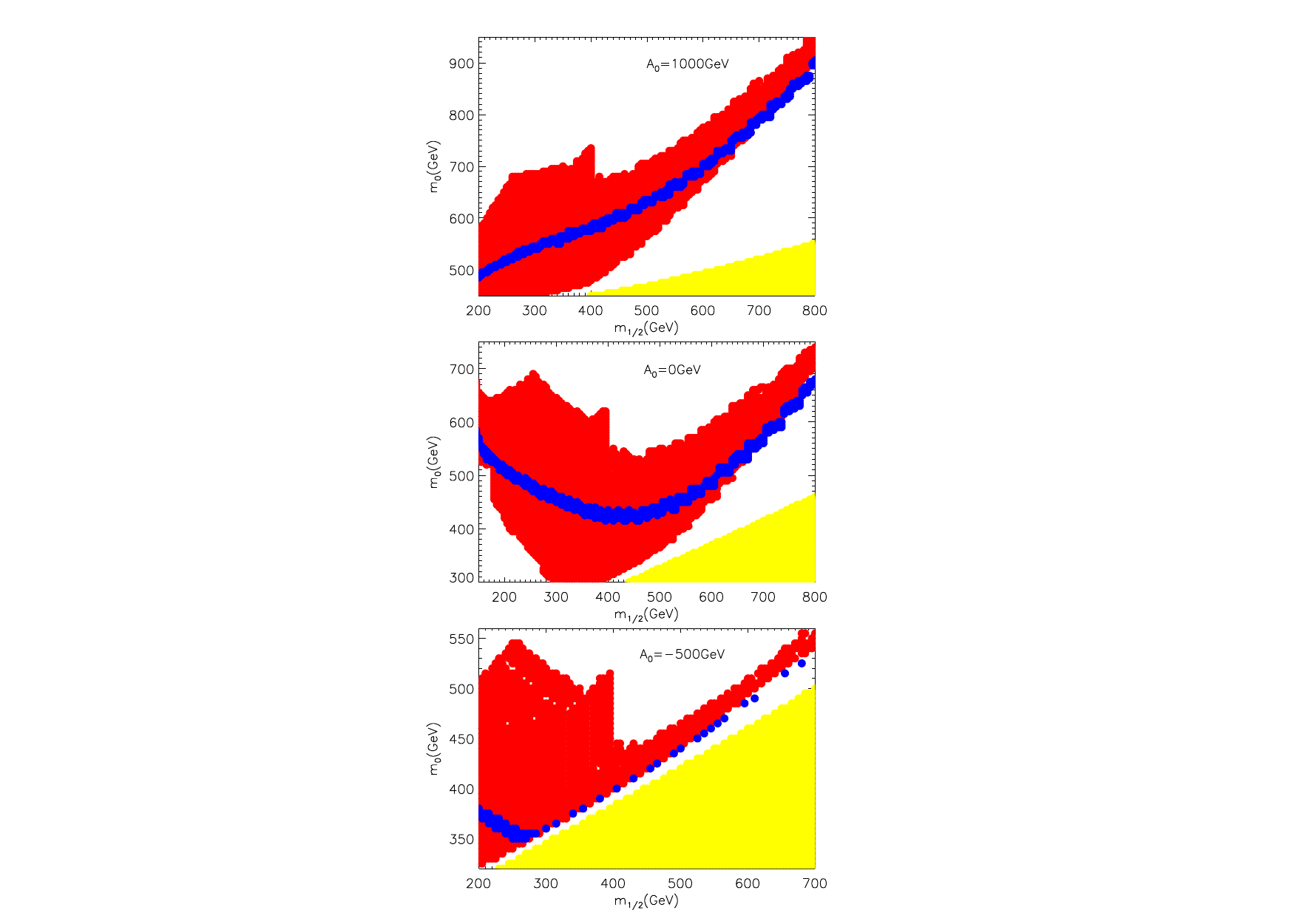}
\caption{The same as Fig.~(\ref{tb10new}), but for tan$\beta=50$, and now $A_0=-500$ GeV in the
bottom panel.}
\label{tb50new}
\end{figure}

Repeating the same procedure for larger values of tan$\beta$, it is found
that the intersection region for both criteria becomes larger, but it
gets to be significant for the largest values of this parameter. This
is clearly shown in Fig.~(\ref{tb50new}), which is equivalent to
Fig.~(\ref{tb10new}), but for tan$\beta=50$. In this case the bottom
panel is for $A_0=-500$ GeV. It is clear from the figure that for
these values of tan$\beta$ both criteria are consistent, as shown by the
large intersection area for values of $A_0$ in the interval $[0,1000]$
GeV. For negative values of $A_0$ the intersection region decreases
with $A_0$, see the bottom panel of the figure. For even lower
values of $A_0$ the intersection becomes insignificant.

\begin{figure}\centering
\includegraphics[width=8cm]{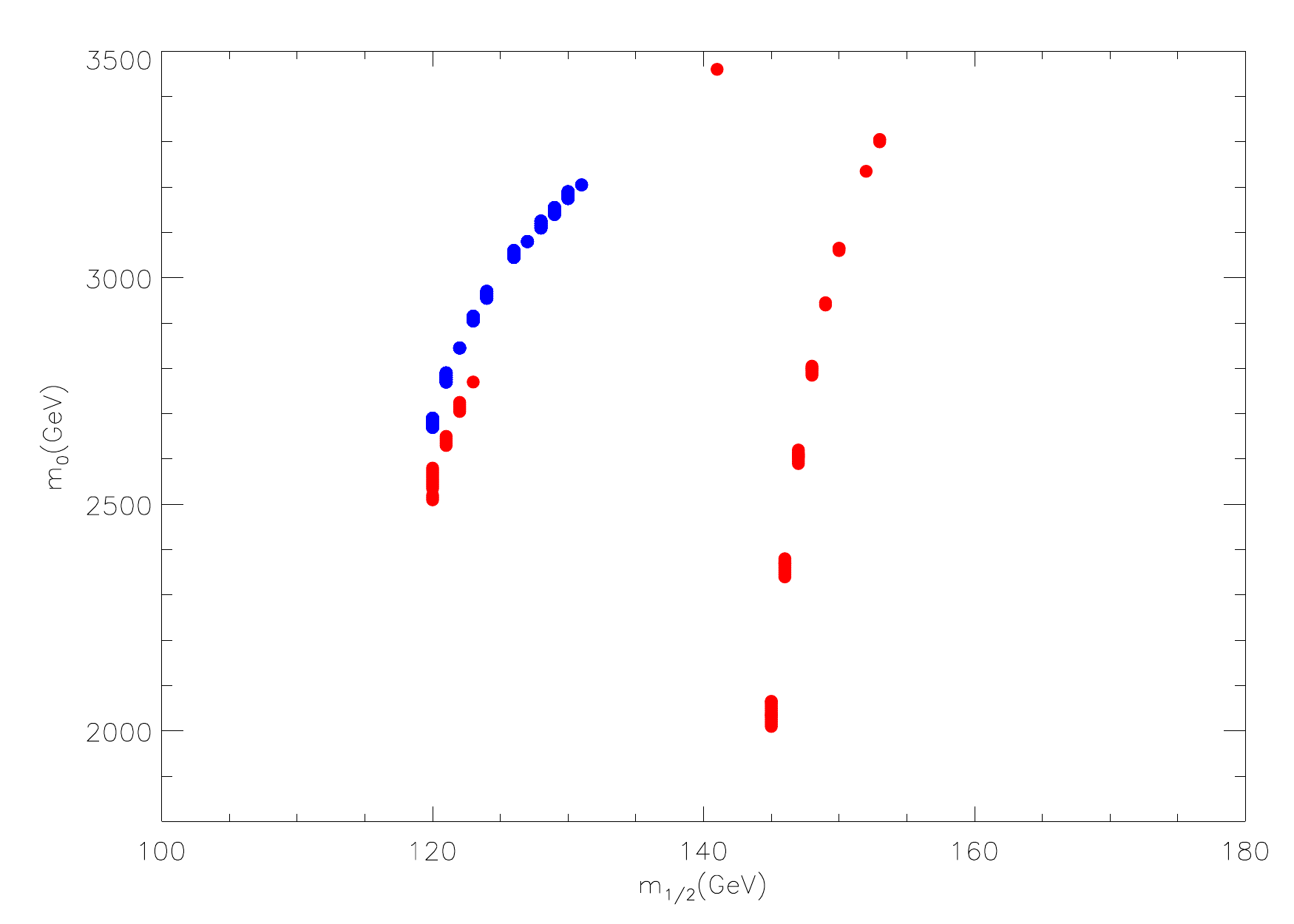}
\includegraphics[width=8cm]{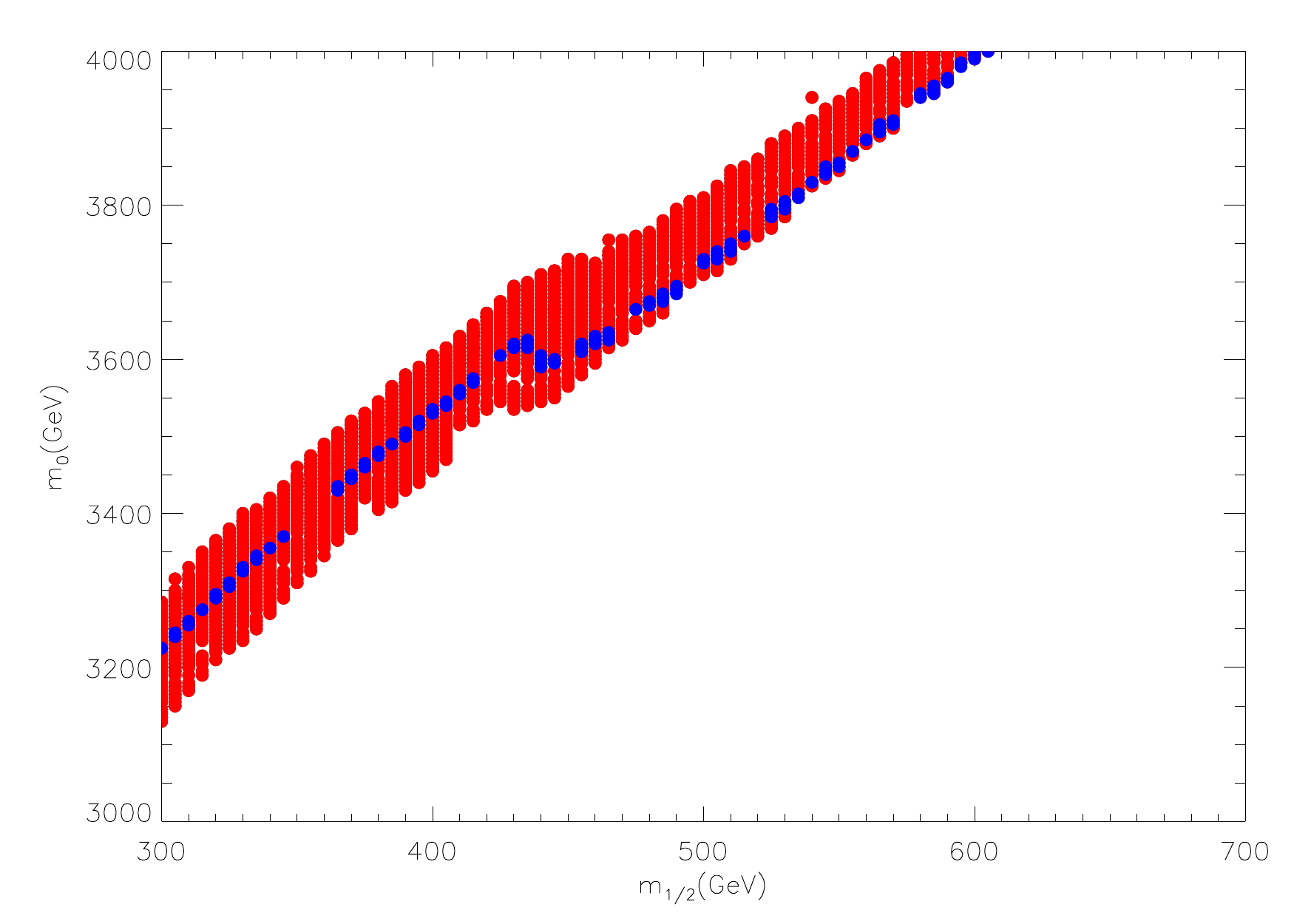}
\caption{Allowed regions in the parameter space for AC (lighter
  gray/red) and EC (darker grey/blue) in the mSUGRA model with
  $A_0=0$, sgn$\mu=+$, tan$\beta=10$, top panel, and tan$\beta=50$, bottom
  panel.  The region shows the so called Focus Point region.}
\label{results2}
\end{figure}

In Figs.~(\ref{results2}) we present the same analysis but for 
the Focus Point region, and for the central value $A_0=0$.  The
situation is consistent with the previous results, both criteria
intersect for $\tan \beta=50$ and there is nearly no intersection for
$\tan \beta=10$.

This analysis allows us to arrive to one of the main results of our
work. The use of both criteria favours large values of tan$\beta$.

\begin{figure}\centering
\includegraphics[width=8cm]{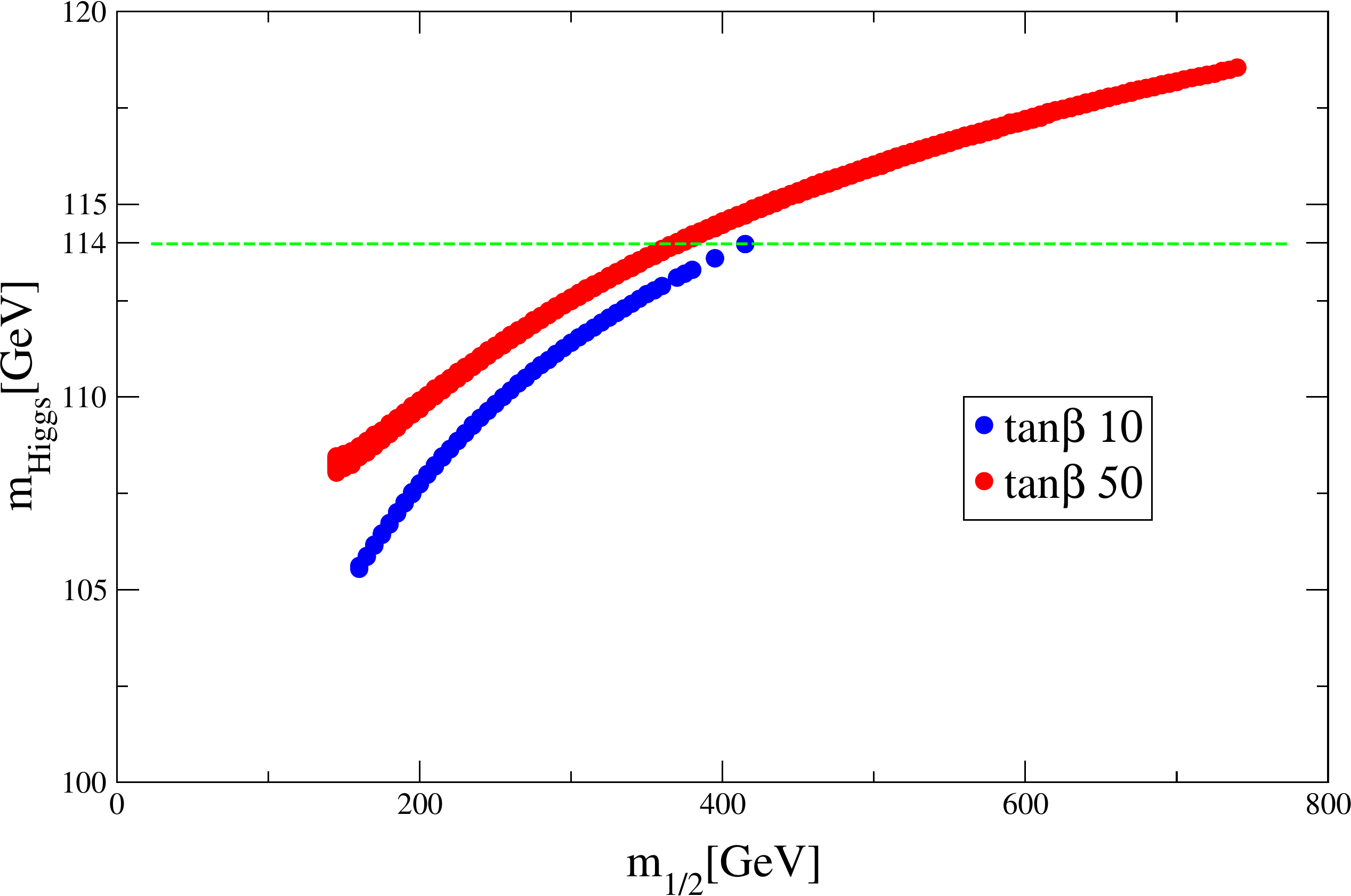}

\caption{Allowed values for $M_{Higgs}$ as function of $m_{1/2}$. 
As can be seen from the figure, the present bound on the Higgs mass
favours a large $\tan \beta$.}
\label{fig:Mhiggsvsm12}
\end{figure}

\begin{figure}\centering
\centerline{\includegraphics[width=8cm,angle=0]{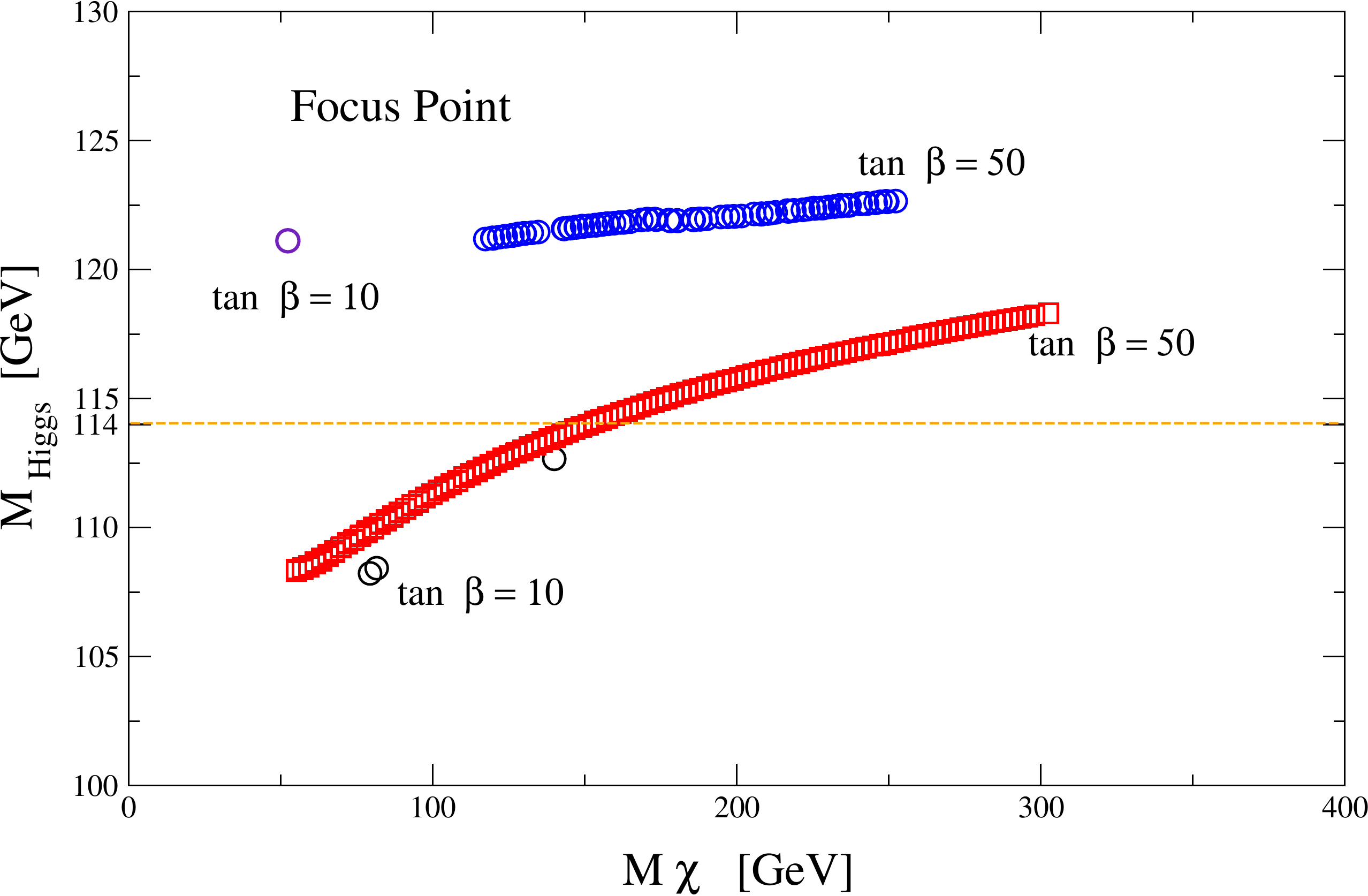}}
\caption{The figure shows the LSP mass plotted versus the Higgs mass,
  points above the dashed line are the allowed values for the LSP. The
points in blue correspond to the Focus Point region, the ones in
red to the bulk and coannihilation regions.}
\label{fig:chi-higgs1}
\end{figure}

In Figs.(\ref{fig:Mhiggsvsm12}) and (\ref{fig:chi-higgs1}) we show the
allowed values for the LSP and the Higgs mass after constraining the
parameter space with the abundance and entropy criteria.  As can be
seen from Fig. (\ref{fig:Mhiggsvsm12}), the current limit for the
Higgs favour, combined with the AC and EC criteria, favours even more
a large value of $\tan \beta$.  This, in turn, puts a constraint on the
allowed SUSY mass spectra of the bulk and coannihilation regions: it
gives an LSP of mass $m_{\chi} \sim 140$ GeV for $\tb ~ 10$, and a lower
bound for the LSP mass $m_{\chi} \gtrsim 150$ GeV for large $\tb$.


 Further analysis, which is currently under way, is
required to give more precise conclusions about this new method to
constrain the parameter space of the mSUGRA model \cite{nuevo}.
\vspace{0.3cm}

 We acknowledge partial support by CONACyT\\ M\'exico, under
grants 32138-E, 34407-E and 42026-F, and PAPIIT-UNAM IN-122002,
IN117803 and \\
IN115207 grants.  JZ acknowledges support from DGEP-UNAM
and CONACyT scholarships.

\end{document}